\documentclass[showpacs,11pt,preprintnumbers,amsmath,amssymb]{revtex4-2}
\usepackage{graphicx}
\usepackage{color}
\usepackage{appendix}
\usepackage[english]{babel}
\usepackage{color}
\usepackage{amsmath, amssymb}
\usepackage{braket} 
\usepackage{subcaption} 
\usepackage{caption}
\usepackage[utf8]{inputenc}  
\usepackage[T1]{fontenc} 
\usepackage{soul}

\usepackage[colorlinks=true,linkcolor=blue,citecolor=blue,urlcolor=blue]{hyperref}
\makeatletter

\begin{document}

\title{Dynamic Synaptic Modulation of LMG Qubits populations in a Bio-Inspired Quantum Brain}

\author{J. J. Torres$^{1,3,*}$ and E. Romera$^{2,3}$}

\affiliation{$^1$ Departamento de Electromagnetismo y Física de la Materia and 
	Instituto Carlos I de F\'\i sica Te\'orica y Computacional, Universidad de Granada, Fuentenueva s/n, 18071 Granada,
	Spain}
\affiliation{$^2$ Departamento de F\'{\i}sica At\'omica, Molecular y Nuclear and 
	Instituto Carlos I de F\'\i sica Te\'orica y Computacional, Universidad de Granada, Fuentenueva s/n, 18071 Granada,
	Spain}
\affiliation{$^3$ Cátedra del Consejo Social en Energía y Tecnologías Cuánticas, Universidad de Granada, 18071, Granada, España}

\affiliation{{\rm *} \mbox{Corresponding author: J.J. Torres, email: jtorres@onsager.ugr.es}}

\begin{abstract}
We present a biologically inspired quantum neural network that encodes neuronal populations as fully connected qubits governed by the Lipkin-Meshkov-Glick (LMG) quantum Hamiltonian and \textcolor{black}{modulated by a synaptic-efficacy feedback implementing activity-dependent changes in the collective time scale.} The framework links collective quantum many-body modes and \textcolor{black}{collective-state structure} to population homeostasis and rhythmogenesis, outlining scalable computational primitives \textcolor{black}{long-lived operating regimes, activity-dependent oscillation periods, and size-dependent robustness} that position LMG-based architectures as promising blueprints for bio-inspired quantum brains on future quantum hardware.
\end{abstract}

\maketitle
\section{Introduction}

The study of learning and information processing is a central theme in neuroscience, neural networks theory, and artificial intelligence, drawing the attention of the scientific community for decades. Recently, these topics have extended for quantum systems. For instance, advances in quantum computing motivated the creation of autonomous systems properly designed to try to find the quantum advantage in information processing, giving rise to emerging fields such as quantum machine learning and quantum artificial intelligence \cite{Biamonte2017,Dunjko2018}. Within this context, several proposals have described \textcolor{black}{quantum learning methodologies} capable of estimating states or unitaries  \cite{Fischer2000, Manzano2009}, implementing quantum reinforcement learning \cite{Dong2008,Briegel2012,Mautner2015,Chen2020,Lockwood2020,Xiao2022}, and constructing quantum neural networks (QNNs) \cite{Torrontegui2019, Chakraborty2020}. Also, variational quantum algorithms have been developed recently and used for such tasks \cite{Cerezo2021}.

Theoretical models have been proposed for both single quantum neurons \cite{Cao2017QuantumNeuron, Tacchino2019ArtificialNeuron, Kristensen2021SpikingQuantumNeuron}, and network architectures, including quantum perceptrons \cite{Pechal2021QuantumPerceptron, Wiebe2016QuantumPerceptron} and Hopfield networks \cite{Rotondo2018} whose computational properties and storage capacity have been recently reported \cite{Torres2024}. Research in this area aims to determine whether such quantum analogues can surpass classical neural networks in pattern recognition and classification and memory storage capacity, while also incorporating biological inspiration \cite{Torres2022}. Within this scenario, typically in most of the recent works concerning developing of quantum neural networks, binary neurons are replaced by qubits with simplified interactions. However, most of the classical and biologically inspired neural network models emphasize the crucial role of synapses \cite{Amit2012ModelingBrainFunction} -- nonlinear elements responsible for transmitting information through pair wise processes such as neurotransmitter release and recycling and high-order interactions involving astrocyte's control of synaptic transmission \cite{Menesse2025AstrocyteControl}. Thus, experimental neuroscience has shown that synapses are dynamic, activity-dependent mechanisms whose transmission efficiency can either decrease (synaptic depression) or increase (facilitation) with presynaptic activity \cite{Tsodyks1998}. These mechanisms have major computational consequences \cite{Torres2013}, influencing memory capacity \cite{Torres2002, Mejias2009}, dynamic memory formation \cite{Pantic2002,Torres2007}, and stochastic resonances during weak-signal processing \cite{Mejias2011}.
They can also lead to an imbalance between excitation and inhibition, causing a quick explosive increase of excitatory activity that originates intriguing brain waves \cite{Pretel2021} with different information content \cite{Menesse2024}.

Building on these previous insights, recently has been proposed a quantum synapse framework incorporating synaptic plasticity, in which a quantum system with biologically inspired activity-dependent coupling between qubits was analyzed to study the effects of synaptic depression on qubit interactions and entanglement \cite{Torres2022}. However, that model was limited to a small-scale system of only two qubits, which does not adequately represent the behavior of large quantum networks. Furthermore, the detailed treatment of qubit interactions in that study makes the approach less practical for extending to systems with many qubits, as the dimensionality of the state space increases rapidly with system size.


The Lipkin--Meshkov--Glick (LMG) Hamiltonian is a compact yet highly expressive framework for exploring quantum many-body phenomena. Originating in nuclear physics~\cite{Lipkin1965a,Lipkin1965b,Lipkin1965c,RingSchuck1980}, it provides a controlled setting for analyzing particle--particle and particle--hole correlations among neutrons and protons, as well as for benchmarking approximate many-body techniques. Its applicability extends well beyond that context, e.g. in condensed-matter physics it offers an effective description of Bose--Einstein condensates and Josephson-junction dynamics~\cite{MilburnEtAl1997,Leggett2001,MicheliEtAl2003}. Also  in optical physics, it has been used for the metrological quantification of spin-squeezed states and the engineering of multipartite entanglement, both in the presence of an external field and for the quadratic collective-spin Hamiltonian in the absence of a field~\cite{KitagawaUeda1993,JurcevicEtAl2014,RichermeEtAl2014,PhysRevA.108.023722}. Taken together, the LMG model has become a standard tool for modeling collectively interacting two-level quantum systems. Within this framework, the existence of first-, second-, and third-order quantum phase transitions (QPTs) has been demonstrated~\cite{Castanos2006PRB,DusuelVidal2004}. More recently, it has been established that the associated quantum phase diagrams can be rigorously characterized in terms of phase-space delocalization measures, complemented by entanglement-entropy measures~\cite{NagyRomera2012,RomeraCalixtoNagy2012,CalixtoRomeraDelReal2012,CastanosCalixtoPerezBernalRomera2015,CalixtoCastanosRomera2017_JSTAT}. Several recent studies have further highlighted the relevance of the Lipkin--Meshkov--Glick model in contemporary quantum many-body physics and quantum simulation~\cite{Chinnarasu2025,Wang2025,LohrRobles2025,Benzahra2025,Romera2026,Kam2026,Tsypilnikov2026}.
In what follows, we focus on the formulation of the LMG model that arises naturally in one-dimensional lattices of interacting spins -- namely, an anisotropic \(XY\) (Ising-type) model in a transverse field with all-to-all couplings.

With this in mind, we here propose for the first time a quantum neural system biologically inspired including some level of short-term synaptic plasticity in the system and which is based in this LMG model. Such theoretical proposal includes two main features  which should be considered while trying to build a {\em quantum brain}, namely, the system emergent properties represent the collective behavior of a system of $N$ interacting quantum neurons or qubits, and it includes a sort of homeostatic plasticity \textcolor{black}{self-modulating the temporal evolution of the collective excitations} in the system similar to short-term synaptic plasticity present in actual brains \cite{Stevens1995,Markram1996, Abbott1997,Zucker2002,Abbott2004}. 

{\color{black}The motivation for combining these ingredients is to establish a bridge between an effective and scalable description of collective quantum dynamics, as provided by the LMG model, and biologically inspired activity-dependent mechanisms which have important computational consequences for the functioning of actual brains, with the aim of exploring how such feedback processes may regulate and shape the emergent behavior of quantum neural systems or investigate possible  intriguing computational properties of such hybrid system.}

Interestingly, the dimensionality of this system remains computationally manageable, as its cost scales polynomially with system size. This property enables the study of collective behaviors in large quantum neural architectures, in contrast to microscopic quantum models where the Hilbert space dimension grows exponentially with the number of qubits. Moreover, our framework can be viewed as a minimal building block for constructing more complex quantum brain–like systems without a dramatic increase in computational complexity. 

\textcolor{black}{
A possible digital implementation of the model can be formulated as a hybrid quantum--classical circuit protocol, in which the LMG evolution is implemented in a quantum layer, while the activity-dependent synaptic feedback is updated in a classical control layer, as discussed in the Supplementary Information.
}It also provides a platform to explore novel emergent phenomena -- such as quantum memory and learning processes -- with potential applications in quantum machine learning. 

\section{Biologically inspired  quantum brain model}

In this paper, we propose a model for collective oscillations within a \emph{quantum brain} paradigm. Specifically, we describe this {\em quantum brain} as a set of \(N\) quantum neurons represented by qubits, such that an activated neuron corresponds to the excited qubit state \(\ket{1}\) and an inactive neuron corresponds to a qubit in the ground state \(\ket{0}\). To model this {\em brain,} we employ the Lipkin-Meshkov-Glick (LMG) Hamiltonian and incorporate a synaptic biological mechanism akin to that introduced for the interaction of two neurons in \cite{Torres2022}.

By interpreting the quantum states of the LMG model as the quantum states of this brain, we obtain a clear analogy with actual brains, where one similarly defines collective oscillatory states in both healthy (e.g., resting state; up/down cortical states during the wake--sleep transition; task-related collective states) and pathological (e.g., epileptic seizure states) conditions. In this way, it becomes possible to classify the quantum states of the model in a manner analogous to how neuroscientists classify empirical brain rhythms.

We now describe the quantum-brain model. To represent the states of this system we use the LMG Hamiltonian. The model consists of \(N\) mutually interacting two-level systems with infinite-range coupling, i.e., each particle can interact with any other. 
We begin with the intensive Hamiltonian
\begin{equation}
H'=\frac{\varepsilon}{N}\sum_{i=1}^{N}\sigma_i^{z}
+\sum_{i<j}\frac{\gamma_x}{N(N-1)}\,\sigma_i^{x}\sigma_j^{x}
+\sum_{i<j}\frac{\gamma_y}{N(N-1)}\,\sigma_i^{y}\sigma_j^{y},
\label{intensive}
\end{equation}
which models a lattice of $N$ spin-$1/2$ particles with infinite-range (all-to-all) $XY$ interactions, characterized by homogeneous couplings $\gamma_x,\gamma_y$. 
To connect this intensive representation with the collective formulation in terms of total-spin operators $J_\alpha=\tfrac12\sum_{i=1}^{N}\sigma_i^\alpha$, we identify parameters via
$
h=-2\varepsilon/N$,
$g=-(\gamma_x+\gamma_y)/(N-1)$,
$\gamma=(\gamma_x-\gamma_y)/(\gamma_x+\gamma_y)$.

With these definitions, the Hamiltonian assumes the canonical collective form
\begin{equation}
H_{\mathrm{LMG}}
=-\frac{g}{N}\Big[(1+\gamma)\,J_{x}^{2}+(1-\gamma)\,J_{y}^{2}\Big]
-h\,J_{z},
\end{equation}
which is equivalent to \eqref{intensive} up to an inobservable additive constant (zero-point energy). {\color{black}
Since \( [J^{2},H_{\mathrm{LMG}}]=0 \), the Hilbert space decomposes into dynamically invariant sectors of fixed total angular momentum \(j\), and the evolution does not mix subspaces with different \(j\). In the present work, we restrict the dynamics to the collective sector with total angular momentum \(j=N/2\).   This restriction significantly reduces the spectral complexity, as the relevant Hilbert-space dimension is reduced from \(2^N\) to \(N+1\).
} \textcolor{black}{
In the maximally symmetric subspace $j=N/2$, we use the Dicke, or angular-momentum, basis $|j,m\rangle$, with $m=-j,\ldots,j$. Equivalently, we label these states by the number $n$ of excited neuronal qubits, $n=m+j$, and write
}
\[
\textcolor{black}{
|n\rangle \equiv |j=N/2,m=n-N/2\rangle .
}
\]
\textcolor{black}{
The fully silent, exactly symmetric semi-activated, and fully saturated states correspond respectively to
}
\[
\textcolor{black}{
|n=0\rangle,\qquad |n=N/2\rangle,\qquad |n=N\rangle,
}
\]
\textcolor{black}{
for even $N$.
} 

{\color{black}
From a neural perspective, the model can be viewed as describing a population of $N$ two-level quantum neurons, where each unit can occupy an inactive or active state. The interaction terms generate collective transitions between different global activation patterns, effectively redistributing activity across the network without changing the total number of neurons. In this picture, the quadratic spin operators encode cooperative processes that couple the activity of all units, leading to emergent collective dynamics. Additionally, the presence of parity symmetry imposes further constraints on the accessible configurations, structuring the global behavior of the system.
}  Here $h$ is an external field and the parameter $\gamma$ is an anisotropic parameter controlling the lack of continuous rotational symmetry in the $XY$ plane, i.e. the relative weight of pairing terms, \textcolor{black}{while preserving the discrete parity symmetry $(J_x,J_y)\rightarrow(-J_x,-J_y).$
}

To account for dynamical processes affecting qubits interactions similar to those reported in \cite{Torres2022}, we consider $g(t)=g_0r(t)$ where  \(r(t)\) is a dimensionless, time-dependent coupling modulation, and $g_0$ is a maximum coupling constant. Additionally, we assume that the time-dependent coupling modulation \(r(t)\) evolves according to a differential equation inspired by models of short-term synaptic depression, \textcolor{black}{a kind of short-term synaptic plasticity mechanism widely studied in both experimental and theoretical neuroscience literature \cite{Tsodyks1998,Zucker2002, Torres2013}}, characterized by a neurotransmitter recovery time \(\tau_r\) constant and a time-dependent release probability \(U(t)\) to account also for a synaptic facilitation mechanism \textcolor{black}{also involved in short-term synaptic plasticity in actual brains (here, we use \(U(t)\) following the standard notation in the neuroscience literature, and it should not be confused with the quantum time-evolution operator)}. \textcolor{black}{Short-term depression and facilitation are respectively related with the slow recovery (controled by $\tau_r$) of neurostransmitter vesicles near the synaptic cleft in actual synapses which induces a decrease in the postsynaptic response for high frequency incoming presynaptic neural spiking activity, and for the  increase in neurotransmitter release probability induced by accumulation of calcium at the the cytosol of the presynaptic neuron every time an action potential arrives due to the opening of calcium channels and the subsequent influx of calcium from the extracellular medium, an excess of calcium than has been reported to increase the probability of vesicles to be released (see for instance \cite{Bertram1996, JackmanRegehr2017}). The dynamics of these two synaptic mechanisms can be simply expressed as}
\[
\dot{r}(t) = f_r\big(r(t), \langle J_z \rangle, \tau_r, U(t)\big).
\]
with 
\[
\dot{U}(t)=f_U\big( U(t),\langle J_z \rangle, \tau_f, \cal{U}\big)
\]
where $\cal{U}$ is the release probability in absence of synaptic facilitation (see below).
The variable $r(t)$ will represent then the hypothetical level of short-term synaptic plasticity in the {\em quantum brain}.
The modulation \(r(t)\) is dynamically linked to the quantum evolution (via the von Neumann equation), producing a nonlinear feedback term in the dynamics of the quantum brain and to the evolution of the release probability $U(t)$ as follows
\begin{equation}
	\frac{d\rho(t)}{dt} \;=\; -\frac{i}{\hbar}\,\big[H_{LMG}(t),\rho(t)\big],
	\label{eq:vonNeumann}
\end{equation}
\begin{equation}
	\frac{dr(t)}{dt} \;=\; \frac{1 - r(t)}{\tau_r} \;-\; U(t)\,r(t)\; \langle E \rangle_t,
	\label{eq:continuous_r}
\end{equation}
\begin{equation}
	\frac{dU(t)}{dt} \;=\; \frac{{\cal{U}} - U(t)}{\tau_f} \;+\; {\cal{U}}\big(1-U(t)\big)\; \langle E \rangle_t,
	\label{eq:continuous_u}
\end{equation}
where the expectation value \(\langle \cdot \rangle_t\) is taken with respect to \(\rho(t)\).

Here, \(E\) is an operator that encodes a collective property of the quantum-brain model. The choice of \(E\) is quite general, enabling us to study how different observables shape the emergent dynamics. As a first step, and in line with the neuroscience analogy, we consider
\begin{equation}
	E(t)=\frac{1}{2}+\frac{J_z(t)}{N},
\end{equation}
which quantifies the level of collective excitation in the quantum brain. {\color{black}
This quantity admits two equivalent interpretations: as a collective polarization in the spin representation, or as the average activity level of the qubit population in the occupation-number picture.
}

In what follows and for simplicity, we shall work with the quadratic collective-spin Hamiltonian in the absence of an external field, i.e., $h=0$. 
{The extension of the present study for cases with non-zero $h$ can be related with the situation in which our quantum brain is receiving external stimuli from outside of the system as in classical neural networks, or from the senses in actual brains.

\textcolor{black}{For this main case, the synaptic modulation $g(t)=g_0 r(t)$ acts as a scalar
prefactor of the LMG Hamiltonian. Therefore, $r(t)$ does not modify the geometric trajectory
of the quantum state, but rather the rate at which this trajectory is traversed. In this sense,
the feedback {can} be interpreted as an activity-dependent time rescaling {when $h=0,$} affecting oscillation
periods, dwell times, and the temporal occupation of low- and high-activation regions.}

In the next section, we present the main results obtained with this quantum-brain model. \textcolor{black}{
Unless otherwise stated, all quantities throughout the manuscript are expressed in arbitrary units (a.u.) and time will be expressed in natural units so it is typically measured in inverse energy units.
}

\section{Results}
\subsection{Emerging Collective states in the quantum brain model}

\begin{figure}[ht!]
\begin{center}
\includegraphics[width=\linewidth]{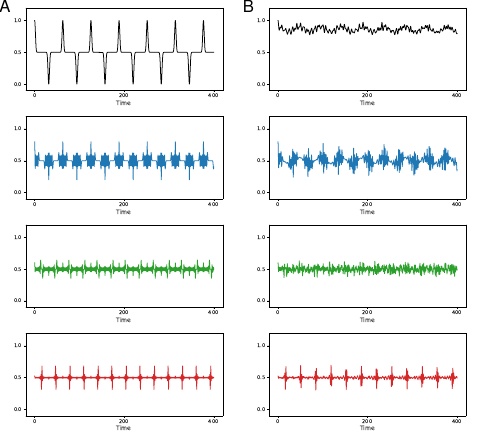}
\end{center}
\caption{Particular dynamical behavior emerging in our quantum brain model, starting from
different initial states with different \textcolor{black}{percentages} of excited neuronal qubits
corresponding from top to bottom to 100\%, 80\%, 60\%, 53\%.
\textcolor{black}{The y-axis represents the collective activation
$E(t)=1/2+\langle J_z\rangle/N$.}
Panel (A) \textcolor{black}{corresponds} to $\gamma = 1$ and (B) to $\gamma = 0.9$,
respectively. Other parameters were $g_0 = 2$, and $\tau_r = 0$, $\tau_f = 0$
or equivalently $r(t)=1$ and $U(t)=U=0.5$ for all $t$.}
\label{figuracolective}
\end{figure}
The first thing we observe when analyzing the system is the very rich dynamical behaviour depending on many dynamical aspects it includes as, for instance, the considered initial quantum state and the different values of system's relevant parameters. The system size also plays a main role since it increases the number of possible quantum states that can be reached during the quantum evolution of the system as we will see below. 

As an initial look at the system’s behaviour, in figure \ref{figuracolective} we illustrate some typical oscillatory behaviour emerging in the system starting from different initial quantum states for some relevant parameters and for the case of $N=40$, \textcolor{black}{ and in the limit $\tau_r\rightarrow 0$ and $\tau_f\rightarrow 0$  which implies $r(t)\rightarrow 1$ and $U(t)\rightarrow {\cal U}=0.5$ (in fact $\tau_r=0$ means not dynamics for $r(t)$ so $r(t)=constant=1=\lim_{\tau_r\rightarrow 0}r(t)$ and similarly $\tau_f=0$ means not dynamics for $U(t)$ so $U(t)=constant={\cal U}=\lim_{\tau_f\rightarrow 0}U(t).$)} For comparison purposes we also consider the cases $\gamma=1$ and $\gamma=0.9.$

Next, \textcolor{black}{we consider the behaviour of the system under the complete dynamics (\ref{eq:vonNeumann}-\ref{eq:continuous_u}) for $\tau_r>0$ and $\tau_f=0$} and  studied in more detail how the dynamical behavior of the system changes when system size and the initial state are varied. Thus, we characterize the dynamics of our {\em quantum brain} architecture under different initial conditions defined over the population state of the qubit neuronal population. \textcolor{black}{
The first observation in this neuronal-qubit system is that, when the initial condition corresponds to an exactly symmetric semi-activation, the collective activation remains strictly stationary at $\langle E(t)\rangle=1/2$, independently of the system size $N$, for $h=0$. This follows from the symmetry of the Hamiltonian under $m\to -m$, which ensures $\langle J_z(t)\rangle=0$ when the initial state is $|j,0\rangle$. Therefore, no oscillations are observed in the population activity.
}

Secondly, we consider an initial state of population semi-activation, in which approximately (but not exactly) half of the neuronal qubits are excited, as illustrated in figure \ref{fig:1}. The temporal evolution of this fraction is denoted by $\langle E(t)\rangle$ . We have explored different network sizes but we illustrate here the cases $N=10$ (left panel) and $N=80$ (right panel) for $\gamma=1$ and $g_0=0.05$. In the figure we display the averaged fraction (or count) of excited neuronal qubits $\langle E(t)\rangle$ as a function of time, alongside the time course of synaptic efficacy $r(t)$. The temporal windows are adjusted to optimize the readability of dynamical regimes ranging from $0$ to $4000$ units for $N=10$, and from $0$ to $10000$ for $N=80$. 

The results of our study reveal a strong anticorrelation between $r(t)$ and $\langle E(t)\rangle$, similar to what is observed in biological systems: when population activity (the number of excited qubits) reaches its maximum, it leads to a reduction in synaptic efficacy, which subsequently reaches a relative minimum. Conversely, when the level of excited qubits is at its minimum, the synaptic efficacy recovers and attains its maximum value.

On average, the system remains near the ($N/2$) operating point, consistent with population-level homeostasis. As $N$ increases, the network exhibits enhanced neurodynamic stability: the variability of $E(t),$ i.e., the oscillation amplitude of the excited fraction, progressively diminishes, indicating more effective gain control and a more confined attractor landscape. Nevertheless, on longer timescales, transient excursions (spikes) in the excited fraction emerge; even so, synaptic efficacy acts as a negative feedback loop that \textcolor{black}{modulates the residence times around the initial population-activation operating regime.}

\begin{figure}[t] 
	\centering
	
	\begin{subfigure}[t]{0.48\textwidth}
		\centering\includegraphics[width=\columnwidth]{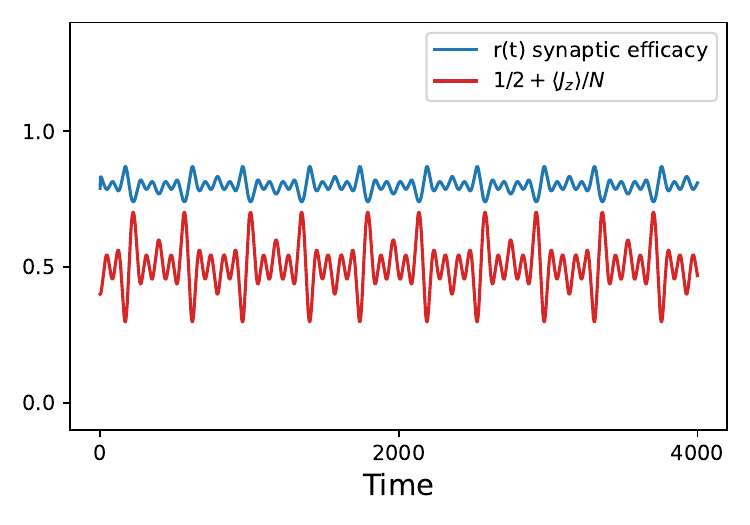}
		\label{fig:da10}
	\end{subfigure}
	\begin{subfigure}[t]{0.48\textwidth}
		\centering\includegraphics[width=\columnwidth]{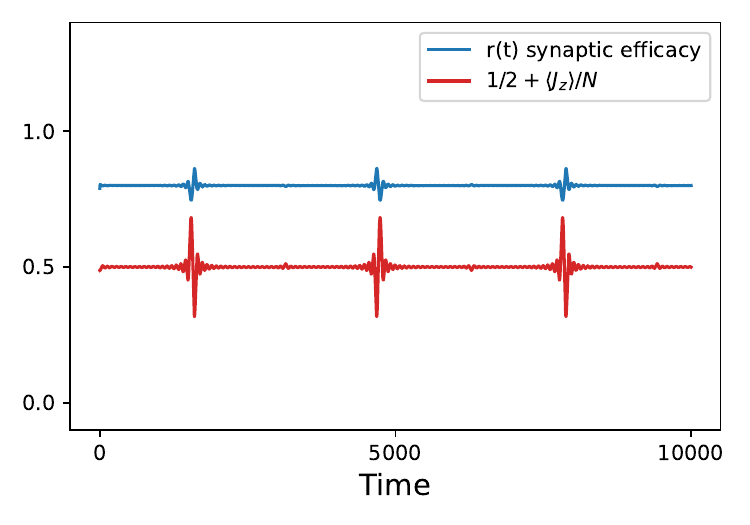}
		\label{fig:dd10}
	\end{subfigure}
	
	\caption{Time evolution of the number of excited neuronal qubits $E(t)=1/2+\langle J_z\rangle/N$ and synaptic efficiency $r(t)$ for 
		(left) $N=10$, and (right) $N=80$ neuronal qubits. Initial state: around half of the neuronal qubits are excited. Other parameters were $\gamma=1,$ $g_0=0.05,$  $\tau_r=1$ \textcolor{black} {and $\tau_f=0$ so one has $U(t)={\cal U}=0.5$ for all time.}}\label{fig:1}
\end{figure}

Third, we consider a fully silent initial condition in which no neuronal qubit is excited (see figure \ref{fig:2}). This scenario is analyzed with the same neuronal-qubit population size as in the preceding case and over identical temporal windows. 
Although the level of synaptic depression is not very high ($\tau_r=1$), the figure suggests that synaptic efficacy can function as a homeostatic feedback mechanism. Initially, relatively \textcolor{black}{high} values of 
$r(t)$ facilitate the activation of excited units when overall activity is low, leading to the growth of the red line toward its maximum. This, in turn, triggers a subsequent decrease in synaptic efficacy (the blue line declining toward its minimum). \textcolor{black}{At this point, $r(t)$ begins to reduce the instantaneous coupling as activity peaks, slowing the subsequent evolution and modifying the activation profile (the red line decreasing after the maximum), while $r(t)$ recovers (the blue line rising after the minimum).}

Moreover, as network size increases, the dwell time within a metastable regime where approximately
half of the neuronal qubits remain excited lengthens. Nevertheless, the dynamics exhibit periodic
excursions toward extreme states fully quiescent (0 \% excited) or fully saturated (100\% excited)
\textcolor{black}{whose residence times are modulated by synaptic homeostasis as explained above,
altering the temporal occupation of these boundary activation regions.}

Finally, it is noteworthy that, starting from an initial condition with no neuronal activity, the temporal evolution of the system spontaneously converges toward a population-level activity pattern consistent with the network  normal  operating regime. 
 \begin{figure}[t] 
 	\centering
 	
 	\begin{subfigure}[t]{0.48\textwidth}
 		\centering\includegraphics[width=\columnwidth]{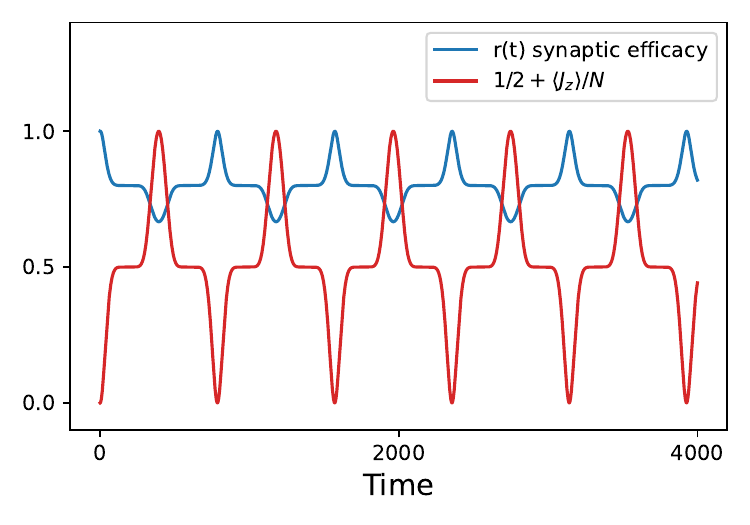}
 		\label{fig:da2}
 	\end{subfigure}\hfill
 	\begin{subfigure}[t]{0.48\textwidth}
 		\centering\includegraphics[width=\columnwidth]{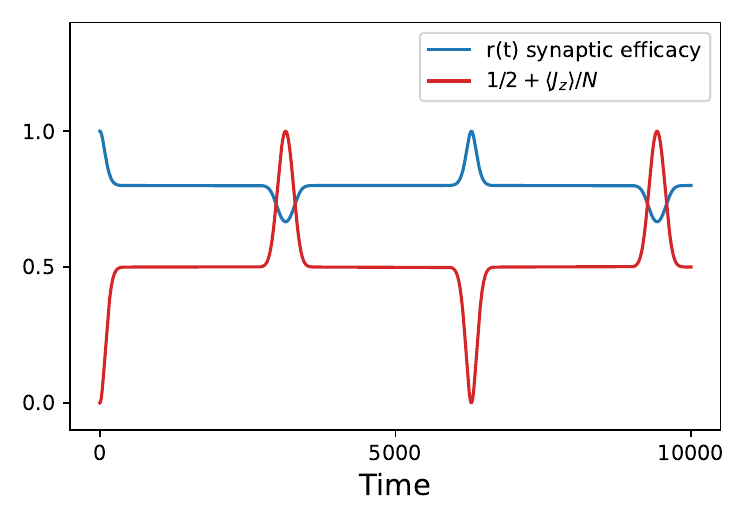}
 		\label{fig:dd2}
 	\end{subfigure}
 	
 	\caption{Time evolution of the number of excited neuronal qubits $E(t)=1/2+\langle J_z\rangle/N$ and synaptic efficiency $r(t)$ for 
 		(left) $N=10$, and (right) $N=80$ neuronal qubits. Initial state: no neuronal qubit is excited. Other parameters were $\gamma=1,$ $g_0=0.05,$ $\tau_r=1$ \textcolor{black} {and $\tau_f=0$ so one has $U(t)={\cal U}=0.5$ for all time.}}\label{fig:2}
 \end{figure}
 
 Lastly, we consider a fully saturated initial state in which all neuronal qubits are excited. The corresponding behavior is illustrated in figure \ref{fig:3}. Once again, synaptic efficacy operates as a homeostatic negative feedback loop: \textcolor{black}{it modulates the effective collective time scale associated with the
hypersaturated initial condition and changes the temporal evolution of the fraction of
excited units.} When the dynamics cross the minimal activity threshold (i.e., enter regions approaching population quiescence), \textcolor{black}{the recovery of synaptic efficacy increases the instantaneous coupling and changes the rate at which the system leaves the low-activity region.}
 
 Consistent with the preceding cases, increasing network size lengthens the intervals of residence around the population operating point near $N/2$ (a metastable regime). Nevertheless, the temporal trajectory exhibits periodic excursions toward boundary attractors: episodes of overshoot with transient elevation to near-total activation, followed by undershoot with de-excitation that can reach complete silencing (0 excited qubits). These cycles of over- and under-activation are accompanied by synaptic homeostasis, \textcolor{black}{which modulates the temporal occupation of different activation regions and keeps the dynamics fluctuating around the network functional operating regime.}

 \begin{figure}[t] 
 	\centering
 	
 	\begin{subfigure}[t]{0.48\textwidth}
 		\centering\includegraphics[width=\linewidth]{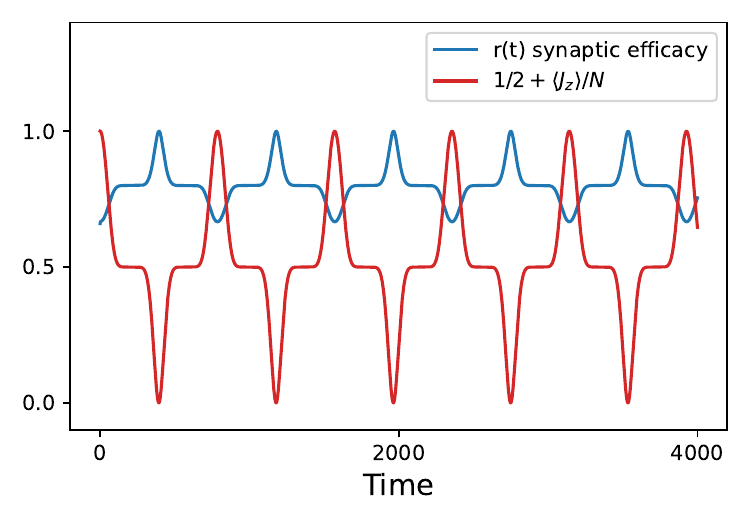}
 		\label{fig:da3}
 	\end{subfigure}\hfill
 	\begin{subfigure}[t]{0.48\textwidth}
 		\centering\includegraphics[width=\linewidth]{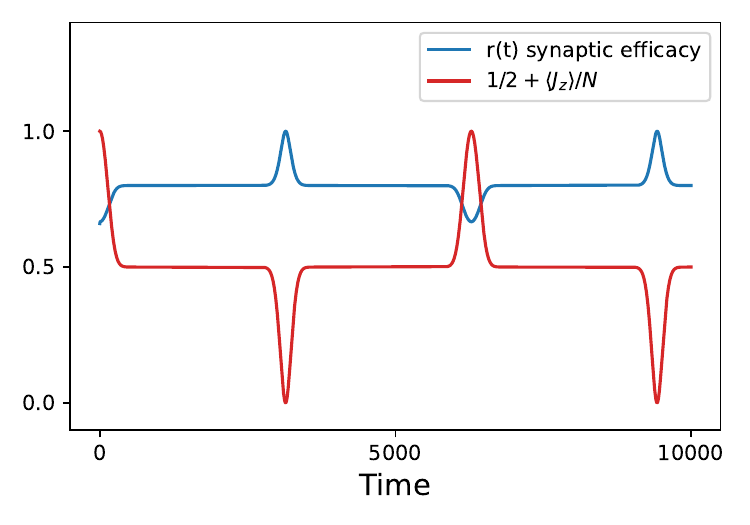}
 		\label{fig:dd3}
 	\end{subfigure}
 	
 	\caption{Time evolution of the number of excited neuronal qubits $E(t)=1/2+\langle J_z\rangle/N$ and synaptic efficiency $r(t)$ for 
 		(left) $N=10$, and (right) $N=80$ neuronal qubits. Initial state:all neuronal qubits are excited. Other parameters were $\gamma=1,$ $g_0=0.05,$ $\tau_r=1$ \textcolor{black} {and $\tau_f=0$ so one has $U(t)={\cal U}=0.5$ for all time.}}\label{fig:3}
 \end{figure}
 
 \textcolor{black}{
To make the comparison across system sizes more transparent, figure~\ref{fig:rescaled_time_comparison} shows $1/2+\langle J_z\rangle/N$ and the synaptic efficacy $r(t)$ as functions of the rescaled time $t/N$, for $N=10,20,40,80$. This representation shows that the main activation and synaptic-recovery events occurs within comparable windows of $t/N$, confirming that the characteristic collective time scale grows approximately with $N$. Nevertheless, the rescaling does not remove all finite-size effects: larger systems display sharper and more localized transitions, whereas for $N=10$ the pulses are broader and the temporal structure is more extended.
}

\begin{figure}[t]
    \centering
    \includegraphics[width=0.32\textwidth]{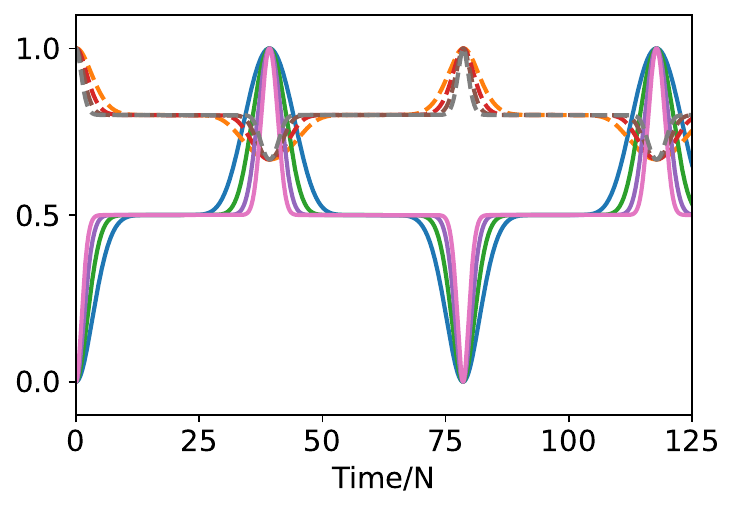}
    \hfill
    \includegraphics[width=0.32\textwidth]{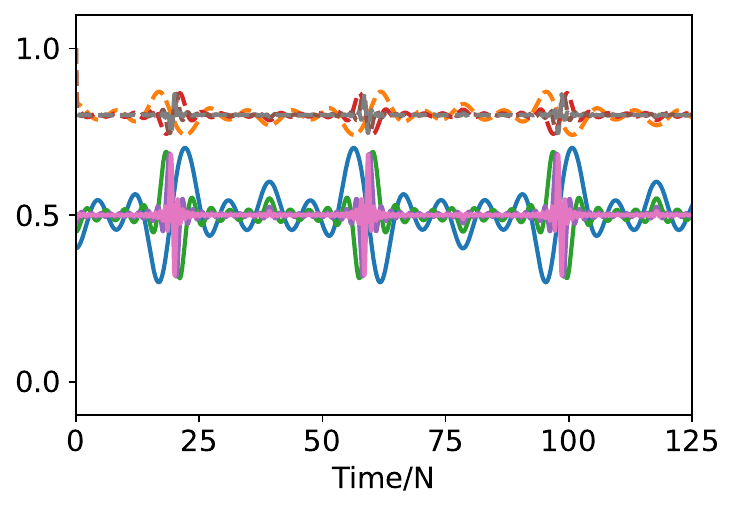}
    \hfill
    \includegraphics[width=0.32\textwidth]{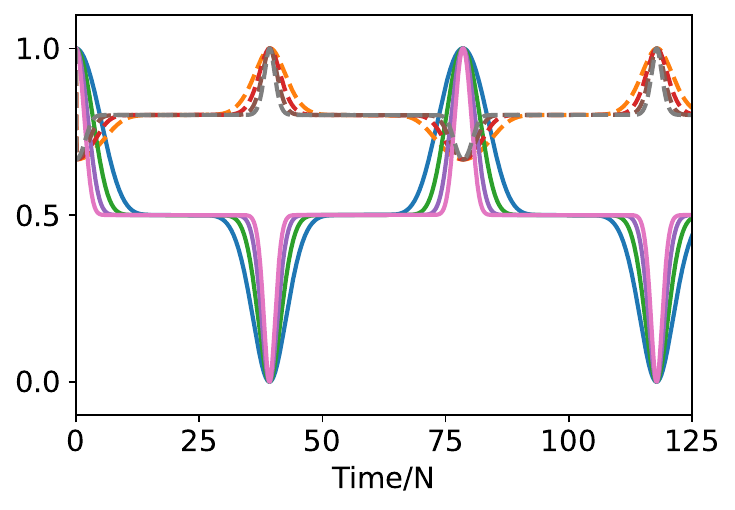}
    \caption{\textcolor{black}{
    Time evolution of the collective activation $E(t)=1/2+\langle J_z\rangle/N$ and the synaptic efficacy $r(t)$ as functions of the rescaled time $t/N$, for different system sizes $N=10,20,40,80$. Solid curves represent $E(t)=1/2+\langle J_z\rangle/N$, whereas dashed curves represent $r(t)$. Solid curves show $E(t)=1/2+\langle J_z\rangle/N$ and dashed curves show $r(t)$. The system sizes are encoded by color pairs: $N=10$ in blue/orange, $N=20$ in green/red, $N=40$ in purple/brown, and $N=80$ in pink/gray. Left panel: fully silent initial condition, with no neuronal qubits excited. Middle panel: approximately semi-activated initial condition, with around half of the neuronal qubits excited, $(N/2-1)$. Right panel: fully saturated initial condition, with all neuronal qubits excited. The remaining parameters are $\gamma=1$, $g_0=0.05$, $\tau_r=1$, $U(t)={\cal U}=0.5$ for all time (i.e. $\tau_f=0$), and $r_0=1$. 
    }}
    \label{fig:rescaled_time_comparison}
\end{figure}

 \begin{figure*}[t] 
 	\centering
 	
 	\begin{subfigure}[t]{0.48\textwidth}
 		\centering\includegraphics[width=\linewidth]{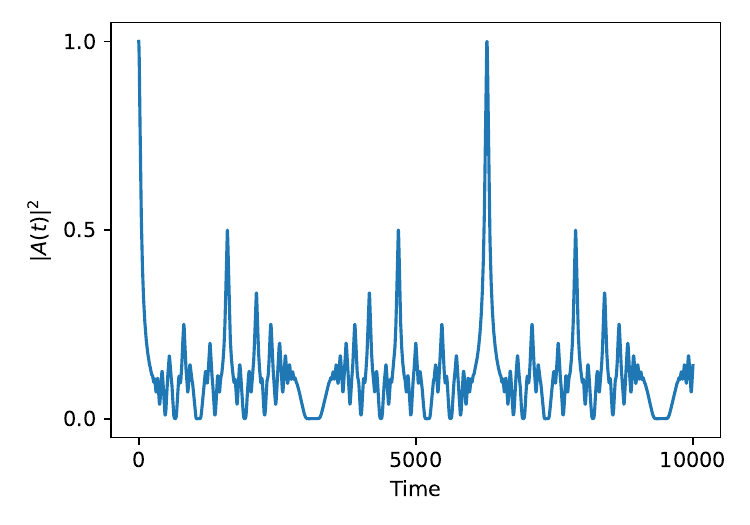}
 		\label{fig:da80}
 	\end{subfigure}\hfill
 	\begin{subfigure}[t]{0.48\textwidth}
 		\centering\includegraphics[width=\linewidth]{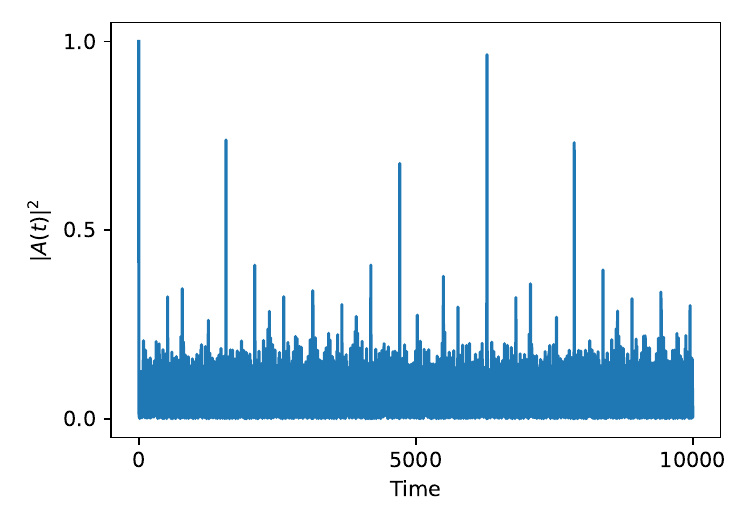}
 		\label{fig:db80}
 	\end{subfigure}\hfill

 	\caption{Time evolution of the fidelity neuronal qubits  for an initial state with
 		(left) no neuronal qubits excited or all the neuronal qubits excited, (right) around half of the neuronal qubits excited ($N/2-1$), with parameters $N=80,$ $\gamma=1$, \textcolor{black} {$\tau_r=1$ with initial synaptic efficacy $r_0\equiv r(0)=1$ and $U(t)={\cal U}=0.5$ (i.e. $\tau_f=0$)}.}\label{fig:fidelidad}
 \end{figure*}
    
      \begin{figure*}[t] 
   	\centering
   	
   	\begin{subfigure}[t]{0.48\textwidth}
   		\centering\includegraphics[width=\linewidth]{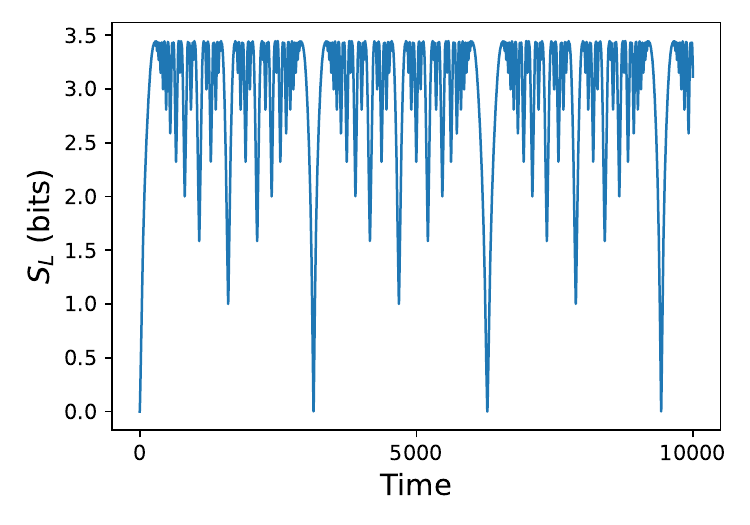}
   		\label{fig:vona80}
   	\end{subfigure}\hfill
   	\begin{subfigure}[t]{0.48\textwidth}
   		\centering\includegraphics[width=\linewidth]{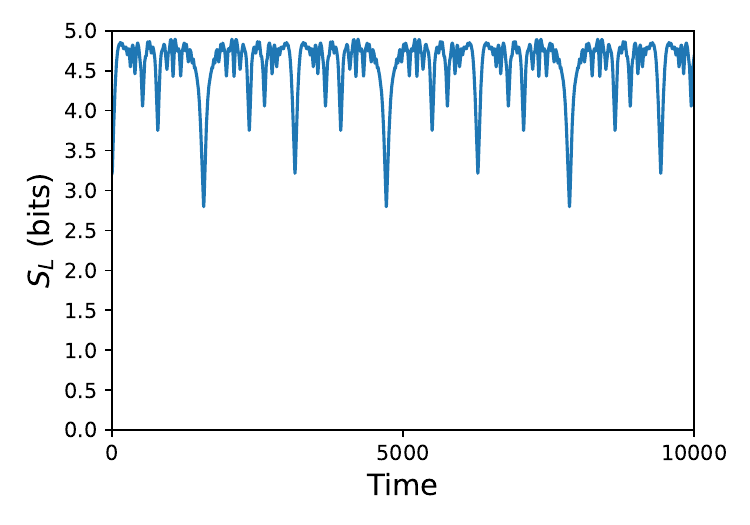}
   		\label{fig:vonm80}
   	\end{subfigure}\hfill
   	
   	\caption{Von Neumann entropy  $S_L$ time evolution of the  neuronal qubits  for an initial state with
   		(left) no neuronal qubits excited or all the neuronal qubits excited, (right) around half of the neuronal qubits excited ($N/2-1$), with parameters $N=80,$ $\gamma=1$, \textcolor{black} {$\tau_r=1$ with initial synaptic efficacy $r_0\equiv r(0)=1$ and $U(t)={\cal U}=0.5$ (i.e. $\tau_f=0$)}.}\label{fig:Von}
   \end{figure*} 
   
  In the figure  \ref{fig:fidelidad} we represent the temporal evolution of the fidelity measured as $|A(t)|^2$
 {\color{black}
\[
|A(t)|^2 = |\langle \psi(0)\mid \psi(t)\rangle|^2 
\]
}with $A(t)$ being the time autocorrelation function. The left panel corresponds to an initial state with all neuronal qubits excited (the behaviour of the system is analogous when one starts with a state with no excited qubits). On the other hand, the right panel shows the dynamics for an initial states with  approximatively half of the qubits being excited. 

In the first scenario, one identifies a temporal structure characterized by a well-defined periodicity, in which the initial state is \textcolor{black}{periodically approached again, as indicated by the revivals of the fidelity}. In the second scenario, the dynamics \textcolor{black}{also shows recurrent departures from and returns toward the initial state}, exhibiting bounded fluctuations with an equally well-defined periodicity.

To quantify the entanglement through the evolution of the system, let us consider a bipartition \(L|(N-L)\) of the Hilbert space and take a block of \(L=\lfloor N/2\rfloor\). \textcolor{black}{Writing the evolved state in the symmetric Dicke basis as}
\[
\textcolor{black}{
|\psi(t)\rangle=\sum_{n=0}^{N} c_n(t)|n\rangle,
}
\]
\textcolor{black}{where \(|n\rangle\) denotes the symmetric state with \(n\) excited neuronal qubits,} the reduced matrix of this block is diagonal in the Dicke basis with probabilities
\begin{equation}
p_k^{(L)} \;=\; \sum_{n=0}^{N} |c_n|^2\,
\frac{\binom{L}{k}\,\binom{N-L}{\,n-k\,}}{\binom{N}{n}},
\qquad k=0,\dots,L,
\end{equation}
in such a way that the von Neumann entropy of the block is given by 
\[
S_L \;=\; -\sum_{k=0}^{L} p_k^{(L)}\,\log_2 p_k^{(L)}.
\]


The figure \ref{fig:Von} shows the time evolution of $S_L$ for $N = 80$, using the same parameters as in the previous simulations. When the initial state has all qubits excited, $S_L$ evolves from $0$ (a pure, non-entangled state) to a maximum close to $3.46$ and periodically returns to values close to zero. {Whenever the system approaches its initial state, the bipartite entanglement between subblocks is transiently suppressed.}

{Conversely, when the initial state has approximately half of the qubits excited, $S_L$ remains at relatively high values throughout the evolution: it starts around $3.3$, increases to reach maximum values of about $4.8$, and only exhibits periodic relative minima slightly below the initial value, never approaching zero. These relative minima coincide in time with the peaks of $E(t)$, indicating brief intervals in which the global state acquires a somewhat simpler structure and the bipartite entanglement between subblocks is partially reduced.}

\begin{figure}[th!]
\includegraphics[width=\linewidth]{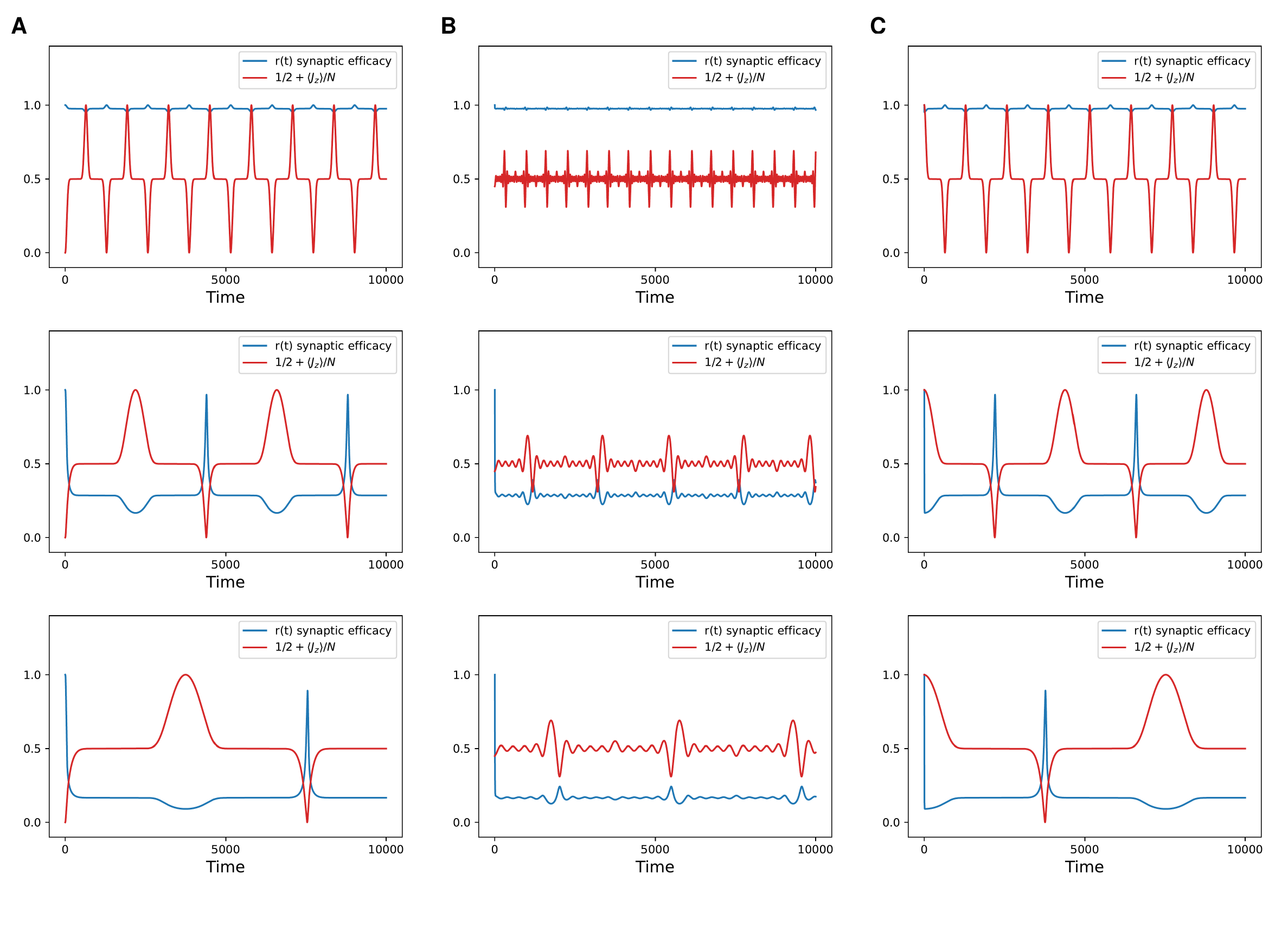}
\caption{Effect of the synaptic plasticity feedback mechanisms on LMG collective dynamics for three different initial states corresponding to (A) no neuronal qubits excited, (B) around half of the neuronal qubits excited ($N/2-1$) and (C) all the neuronal qubits excited. From top to bottom y each panel the parameter $\tau$ controlling the synaptic plasticity feedback has been increased taking the values $\tau_r=0.1, 10, 20.$ Other parameters are $N=20,$ $g_0=0.05,$ $\gamma=1,$ $U(t)={\cal{U}}=0.5$ (i.e. $\tau_f=0$) and $r_0=1.$}
\label{synapticeffectgam1}
\end{figure}

\begin{figure}[th!]
\includegraphics[width=\linewidth]{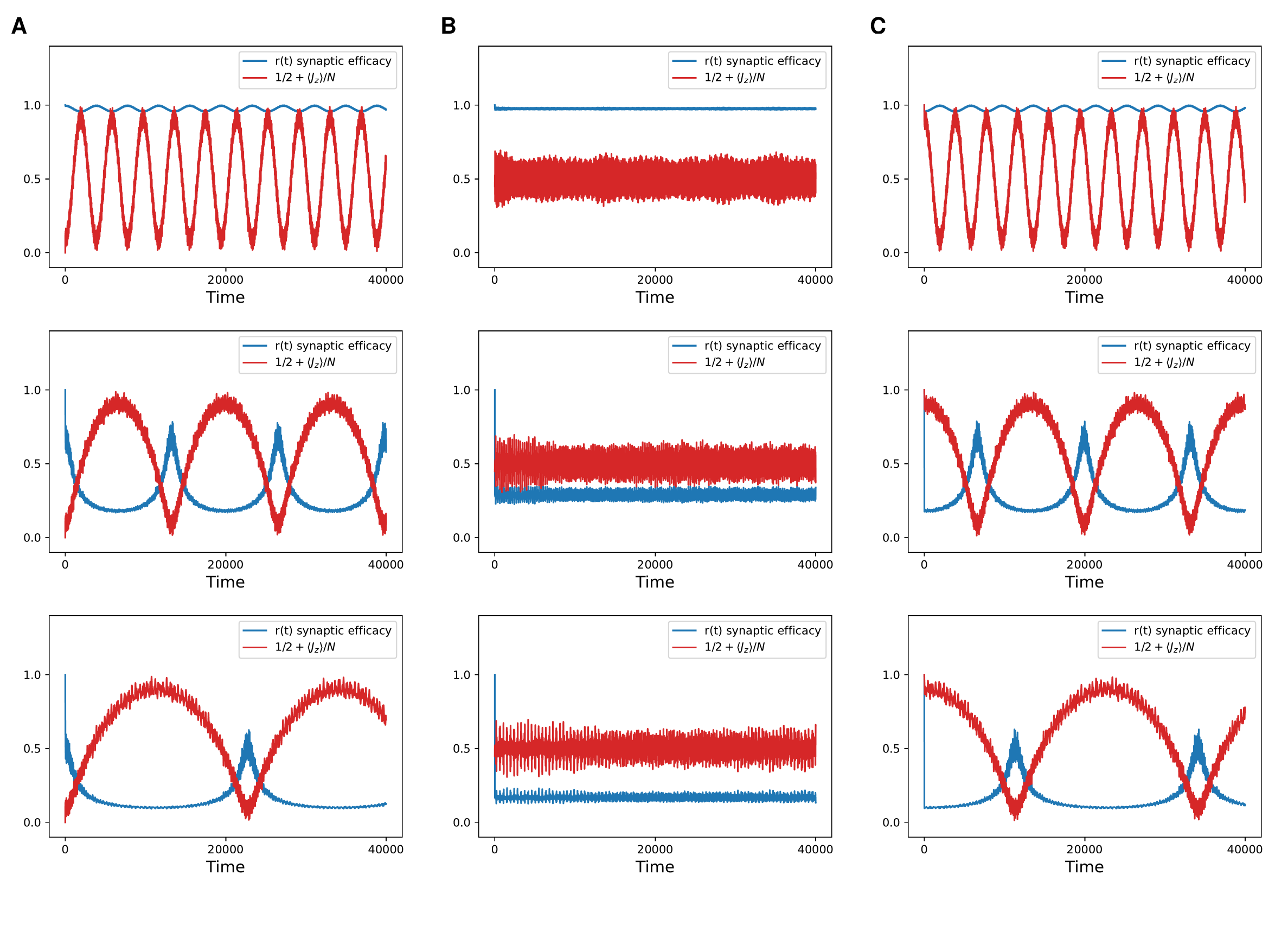}
\caption{Effect of the synaptic plasticity feedback mechanisms on LMG collective dynamics for three different initial states corresponding to (A) no neuronal qubits excited, (B) around half of the neuronal qubits excited ($N/2-1$) and (C) all the neuronal qubits excited. From top to bottom y each panel the parameter $\tau$ controlling the synaptic plasticity feedback has been increased taking the values $\tau_r=0.1, 10, 20.$ Other parameters are $N=20,$ $g_0=0.5,$ $\gamma=0.8,$ $U(t)={\cal{U}}=0.5$ (i.e. $\tau_f=0$) and $r_0=1.$}
\label{synapticeffect}
\end{figure}

\subsection{The effect of synaptic plasticity on collective states}
We have next studied the effect of the synaptic feedback in Eq. \ref{eq:continuous_r} on the features of the emergent  collective dynamics of our {\em quantum brain} system. \textcolor{black}{At the beginning we are going to consider only the effect of synaptic depression dynamics \cite{Tsodyks1998,Pantic2002, Torres2002}, i.e. only Eqs. (\ref{eq:vonNeumann}-\ref{eq:continuous_r}) in our quantum system. From a mathematical standpoint, this mechanism introduces a self-limiting feedback where the synaptic coupling strength decreases in response to high neuronal pre-synaptic activity, representing the depletion of available resources in actual neural systems, or in our quantum system due to the increase of active qubit population, as can be easily see in Eq. (\ref{eq:continuous_r}). As introduced in Section II the model can be easily extended for more complex dynamics including synaptic facilitation \cite{JackmanRegehr2017, Torres2007, Mejias2009} resulting in the  complete system (\ref{eq:vonNeumann}-\ref{eq:continuous_u}). The facilitation mechanism produces an enhancement of the synaptic coupling strength induced also by presynaptic neuron activity in actual neural systems, or by the increase of active qubit population in our LMG quantum brain model as can be easily see in Eq. (\ref{eq:continuous_u}). The results or our study are summarized in figures  \ref{synapticeffectgam1} and \ref{synapticeffect} respectively for \textcolor{black}{one-axis anisotropic case}  as in previous figures (i.e. $\gamma=1$) and for 
\textcolor{black}{a more general anisotropic situation} ($\gamma = 0.8$),
which induces an asymmetry in the terms appearing in the LMG hamiltonian. 
Both figures clearly illustrate the effect of the synapse feedback on the emergent properties of the LMG system.} We also compared the effect of quantum synaptic plasticity when one starts in the same three initial states considered in the previous analysis and that are illustrated, respectively, on panels (A) for an \textcolor{black}{initial state with no excited qubits or low-activation state, panel (B) for a intermediate state with half population of qubits excited or semi-activation state, and (C) for an initial state with all qubits excited or high-activation sate.
Figure \ref{synapticeffectgam1} clearly illustrates that, in the  \textcolor{black}{one-axis anisotropic case} ($\gamma=1$), synaptic feedback creates a non-trivial imbalance during the system's evolution between low- and high-activation states. Specifically, as synaptic depression increases, the system tends to persist in the high-activation state for most of the time. Note that this temporal modification of the state occupation probability is not uniform across all time steps -- as would be expected from a mere effective temperature parameter affecting the LMG Hamiltonian -- since it affects low- and high-activation states differently (note the very fast variation of the low activated sates). This behavior is solely a consequence of the nonlinear dynamics defining the synaptic feedback in our LMG brain model as described in Eqs (\ref{eq:continuous_r}-\ref{eq:continuous_u}).} Similar behaviour occurs for the anisotropic situation ($\gamma\neq 1$) but the situation now is a bit complex. In this case, when the level of synaptic feedback is low (top panels in all cases corresponding to $\tau=0.1$), i.e. $r(t)\approx 1,$ one observes at short time scales that the activity of the system is correlated with the corresponding initial state presenting very high frequency oscillations corresponding {to collective many-body Rabi oscillation}. {However at relatively large time scale (low frequency) the system additionally shows symmetric up/down collective oscillations similar to Rabi oscillations of a single qubit dynamics}. Such oscillations have a typical frequency of $2.5\times10^{-4}$ oscillations per unit time for the set of parameters considered. The oscillatory behavior is symmetric between up and down temporal phases (that is the lower and larger energy states are populated identically during the oscillations) and depends on the parameters of the LMG models such as $\gamma,g_0$ and $N$ (in this case we are considering $N=20$). In fact it is a property of the LMG model. When synaptic feedback start to be important which is produced by increasing the value of $\tau_r$ -- middle and bottom panels -- there is a clear coupling between the intrinsic dynamics of the LMG system and the synaptic plasticity feedback (also visible in Figure \ref{synapticeffectgam1}). The main consequence of such interplay is that the shape of the oscillations dramatically changes in different ways. First, the frequency of the oscillations strongly decreases with $\tau_r$ and secondly a strong imbalance in the population of the \textcolor{black}{low-activation and high-activation states} of the LMG occurs (as already shown in Figure \ref{synapticeffectgam1}), in such a way that the synaptic feedback favors the population of the  high-energy levels when one starts initially at the low energy and at the high energy levels. Interestingly, when one starts a the half populated energy level the system does not show clear low frequency collective oscillations independently of the level of synaptic feedback present in the system -- see panel B -- and synaptic feedback only affect the high-frequency collective many-body Rabi oscillation.  The emerging imbalance in the population of the LMG model induced by the synaptic feedback in the other cases is more strong when $\tau_r$ is larger.

To obtain a clearer picture of how synaptic depression affects the behavior of the collective states in the present quantum-brain model, we computed the evolution of the \textcolor{black}{ von Neumann entropy} $S_L$ for different values of $\tau_r$. Figure~\ref{fig:SL_depresion} compares the temporal evolution of the \textcolor{black}{ von Neumann entropy} $S_L(t)$, 
and the corresponding power spectra for three increasing levels of synaptic depression and when synaptic facilitation is not present (i.e. $U(t)={\cal U}=0.5 $ and $\tau_f=0.$). In the absence of depression, i.e. $\tau_r=0$ (top row), $S_L(t)$ displays fast and highly regular oscillations, alternating between low-entropy episodes associated with the extrema of the collective spin dynamics and high-entropy intervals in the transition regions among these low-entropy states. {
When depression is introduced, i.e. $\tau_r>0$ (middle and bottom rows), a pronounced asymmetry emerges between high- and low-entropy levels, indicative of an imbalance in the transition periods between \textcolor{black}{low-activation and high-activation states}, and vice versa, which in turn results in different time scales for the intervals of high entanglement. This conclusion is corroborated by analyzing in parallel the power spectrum of $S_L(t)$, which, in the absence of depression, exhibits a dominant, narrow, high-frequency peak associated with a well-defined frequency, a clear periodicity and, ultimately, a single characteristic time scale for the high-entropy intervals. By contrast, as depression increases, this peak shifts to lower frequencies and additional harmonics emerge. Furthermore, the main maximum broadens, revealing a progressive blurring of the dominant entropy oscillation and a larger dispersion of the characteristic frequencies. This behavior is consistent with the emergence of longer entanglement or transition periods (in particular during the passage from \textcolor{black}{low-activation to high-activation states}) and with a less sinusoidal signal, compatible with a persistent population imbalance between the extreme collective states.

Taken together, these results indicate that synaptic depression generally slows down and blurs the near-periodic dynamics of the system and, on average, favors more highly entangled states.
\\
\textcolor{black}{To further validate these previous findings, we analyzed the temporal persistence of entanglement by calculating the time intervals during which the Von Neumann entropy exceeds a predefined threshold. This threshold represents a regime of high entanglement for any given value of the synaptic feedback parameter, $\tau_r$. As shown in Supplementary figure 3 of the Supplementary Information, we computed the mean duration of these windows across the entire time series. The results reveal a clear monotonic increase in the average window length as a function of $\tau_r$. This trend confirms that larger synaptic feedback recovering times sustain high entanglement states for significantly longer periods during the system's evolution.}

\begin{figure*}[ht!]
    \centering
    \includegraphics[width=\textwidth]{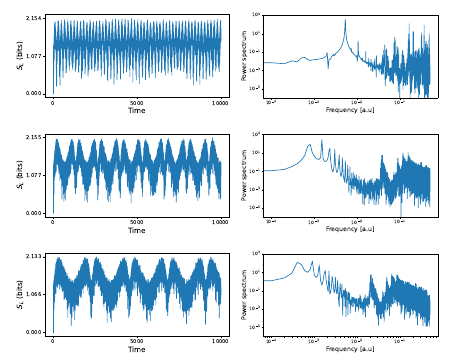}
    \caption{
Temporal evolution of the von Neumann entropy $S_L(t)$ (left panel), and power spectrum of $S_L(t)$ (right panel) for three increasing levels of synaptic depression \textcolor{black}{being $\tau_r=0, 10, 20$ from top to bottom, respectively.
In the absence of depression, }$S_L$ periodically and quickly alternates between low- and high-entropy regimes, but being most of the time between these two states, resulting in an effectively unimodal distribution at at relatively large intermediate value of the entropy. The corresponding spectrum shows a narrow peak at high frequency. As depression increases, the distribution shifts toward higher entropies emerging a second high-entropy peak. The corresponding spectrum shows a peak that moves to lower frequencies with additional secondary harmonics, as depresion increasing indicating longer entanglement periods and increased dwell times in highly entangled states. \textcolor{black}{Other parameters were $N=20,$ $g_0=5,$ $\gamma=0.8,$ $U(t)={\cal U}=0.5$ for all time (i.e. $\tau_f=0$) and $r_0=1.$}}
    \label{fig:SL_depresion}
\end{figure*}

\section{The role of synaptic facilitation}
\begin{figure}[ht!]
\begin{center}
\includegraphics[width=\linewidth]{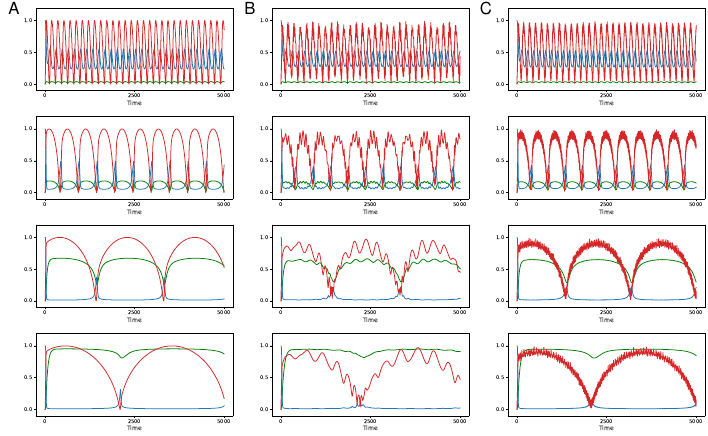}
\end{center}
\caption{The effect of synaptic facilitation on the behaviour of the LMG brain model. The figure shows the cases of system sizes of $N=2$ (A) $N=10$ (B) and $N=20$ (C) for $\tau_f=1,10,100,1000$ from top to bottom respectively. For each $N$ we consider a particular choice for $g_0$ in order to recover the same temporal scale in all cases, so one has $g_0=0.125$ (A), $g_0=1.43$ (B) and $g_0=30$ (C). In each panel it is plotted the level of excitation in the system measured in terms of $E(t)=1/2+\langle J_z\rangle/N $(red line), the level of synaptic feedback measured with $r(t)$ (blue line) and the release probability $U(t)$ (green line). One can observe that independently of the system size, the effect of synaptic feedback and facilitation  is the same after properly rescaling the coupling $g_0.$ Other parameter were $\gamma=0.8$ $\tau_r=100$ and ${\cal{U}}=0.02$ and $r_0=1.$}
\label{figurafacilita}
\end{figure}

We have also explored the role that synaptic facilitation has on the emergence of collective spin states in our quantum brain model. The main findings are summarized in Figure \ref{figurafacilita}. In this analysis we take three different values of  system size $N=2$ (panel A), $10$ (panel B), and $20$ (panel C). Moreover,  we used a small value of the parameter ${\cal{U}}=0.02$ compared with the previous case of synaptic depression in order to see the effect of the synaptic facilitation whose main consequence at the biological level is to increase the release probability $U(t)$ as given by Eq. \ref{eq:continuous_u} during a time scale of $\tau_f$. We also consider a different value of $g_0$ for any value of the system size to have the same temporal scale for population state oscillations and to have a better comparison analysis. Then we increase the facilitation time constant $\tau_f$ for any $N$ depicted in panels A, B, and C from top to bottom.  Note that $\tau_f\approx 0$ means that only synaptic depression is present in the system, which in this case since we have an small value of ${\cal{U}}=0.02$ will correspond to a relatively large level of synaptic depression feedback with $r(t)$ below $0.5$ -- see blue line in the top panel of A, B and C. Also in this situation facilitation of the release probability never occurs so one has $U(t)\approx {\cal{U}}$ (see solid green line in all top panels). When one increases $\tau_f$ (time series from top to bottom in panels A, B and C) we are increasing the time period at which $U(t)$ increases (facilitates). This makes the release probability to increases until it reaches a value $U(t)\approx 1$ corresponding to $r(t)\approx 0$ during any period of large excitation in the system as depicted in the bottom panels. Note that a large value of $U(t)$ also implies an strong depression effect -- see Eq. \ref{eq:continuous_r}. The main consequence is that increasing $\tau_f$ helps to increase the imbalance between \textcolor{black}{low-activation and high-activation collective states}. Moreover facilitation also induces {a systematic imbalance in the transition periods between \textcolor{black}{low-activation and high-activation collective states} and vice versa} (see collective oscillations in red color in the middle panels in A, B, C). Our study also illustrated that, independently of the system size, the effect of facilitation is more or less the same after properly chosen other parameters as $g_0$ in this case. This is not surprisingly since in the LMG hamiltonian we have a global $N$ rescaling $g_0.$ 

\section{Conclusions}
In this work we have presented a theoretical framework to develop a {\em quantum brain} paradigm. Such framework is based in the well known LMG model for the study of qubit population states together with a synaptic homeostatic feedback motivated from neuroscience. In this way, we can build and study a quantum system that show qubits collective states whose dynamics is controlled by the level of synaptic feedback present in the system. This approach {aims to reproduce}, at the quantum level, collective activity states similar to those observed in actual brains. 

Our study shows that the quantum brain model {introduced here exhibits} different homeostatic population dynamics, in such a way that if it starts quiescent or saturated, activity quickly converges to a regime with \(\approx N/2\) excited qubits that acts as a metastable attractor. Larger networks stabilise this operating point, reducing fluctuations of the excited fraction \(E(t)\) and increasing the time spent near it. Similarly to what occurs in actual neural systems, synaptic efficacy in our quantum system provides a negative-feedback loop that \textcolor{black}{reduces the instantaneous coupling during high-activity episodes and increases it as activity decreases, thereby modulating the residence times in different activation regions.}

{
From a quantum perspective, the dynamics depend strongly on the initial state. For ordered configurations (all qubits excited or none), the fidelity attains maximum values equal to one, at which the global state periodically recovers its initial form. The bipartite von Neumann entropy \(S_L\) rises from zero to a maximum and returns to values close to zero when the fidelity reaches unity, indicating a transient suppression of bipartite entanglement. In contrast, for semi-activated initial states (with \(\approx N/2\) excited qubits), \(S_L\) remains high and its minima never reach zero. These minima coincide with the peaks of \(E(t)\), signaling brief episodes in which the system passes close to the initial state, accompanied only by a partial reduction of bipartite entanglement.
Taken together, the model combines homeostatic regulation with quantum entanglement features that are strongly determined by the initial preparation.
}

{Synaptic feedback in the LMG-based “quantum brain” model reconfigures collective behavior in a way that is very similar to synaptic plasticity in biological circuits. Under weak feedback (small \(\tau_r\), \(r(t)\approx 1\)), the system displays high-frequency many-body Rabi oscillations constrained by the initial state and slower, symmetric oscillations between low- and high-energy sectors reflecting the bare LMG dynamics. {As synaptic feedback strengthens (large $\tau_r$), plasticity and intrinsic dynamics become tightly coupled, lowering oscillation frequencies and introducing a systematic imbalance in energy-level occupation, with a growing bias toward high-energy states.}
 The semi-activated initial condition acts as a special operating regime: it does not generate low-frequency collective oscillations for any \(\tau_r\), and feedback only modulates the high-frequency Rabi cycles. Thus, quantum synaptic plasticity, without extra assumptions, plays a role analogous to biological synaptic regulation by tuning time scales and collective patterns and reorganizing, in a state-dependent way, the effective distribution of activity across energy levels via the parameter \(\tau_r\).}

We also observed that synaptic feedback strongly affect collective states entanglement in such a way that the periods of maximum entanglement corresponding to the transitions from a state of non-excited qubits (the low-activation state) to a state of all excited qubits (the high-activation state) enlarges. Also it introduces complexity in the system which is depicted by the emergence of non-periodicities in the time series of the linear entropy.

Our study also shows that synaptic facilitation strongly influences the emergence and time structure of collective states dynamics in the quantum brain model. When facilitation is weak, the dynamics are dominated by synaptic depression, leading to more time symmetric and balanced transitions between activity states. As facilitation becomes stronger, the system exhibits prolonged periods of heightened release probability, which amplify depressive effects and create a stronger separation between excited and non-excited collective states. This also introduces a noticeable asymmetry in the transitions between these states. {Overall, synaptic facilitation enhances the contrast in collective dynamics and also amplifies a systematic imbalance in collective state transitions, and these effects remain consistent across different system sizes when the model is appropriately rescaled.}

The observed imbalance in the levels of the LMG model induced by the synaptic feedback could have interesting applications as for example the use of the LMG with this synaptic feedback for the design of  quantum working memories similar to those already studied in neuroscience \cite{Mongillo2008}. Variants of the present study could also include coupling of LMG systems with different synaptic feedback levels, each one favoring a particular level of quantum qubit populations.

Summing up, the present work represents an initial step toward constructing quantum formulations of biologically inspired neural systems. By doing so, we aim to uncover the potential computational consequences that may arise at the quantum level, particularly those linked to mechanisms that could provide a quantum advantage within biological contexts. This framework opens the door to exploring how quantum effects might enrich or transform the functional principles traditionally associated with neural computation. Additionally, \textcolor{black}{
 the proposed model admits a natural hybrid quantum--classical digital realization, described schematically in the Supplementary Information, where the nonlinear synaptic feedback is implemented through measurement-based classical control between consecutive quantum evolutions.
}

\section{Acknowledgments}
J.J.T. acknowledges support from Grant No. PID2023-149174NB-I00 financed by the Spanish Ministry and Agencia Estatal de Investigación MICIU/AEI/10.13039/501100011033 and ERDF funds (European Union). E.R. acknowledges support from PAIDI group FQM-420 of the University of Granada. Funding for open access charge: Universidad de Granada.

\section{Author contributions}
J.J.T. and E.R. equally contributed to the conception, design, and research of the methods presented in this article. All authors equally contributed to the writing of the manuscript.
\section{Competing interests}
The authors declare no competing interests.
\bibliographystyle{plain} 
\bibliography{references_v4}

\newpage

\begingroup

\setcounter{section}{0}
\setcounter{figure}{0}
\setcounter{table}{0}
\setcounter{equation}{0}
\setcounter{page}{1} 

\renewcommand{\thesection}{S\arabic{section}}
\renewcommand{\thefigure}{S\arabic{figure}}
\renewcommand{\thetable}{S\arabic{table}}
\renewcommand{\theequation}{S\arabic{equation}}
\renewcommand{\thepage}{S\arabic{page}} 

\begin{center}
    \vspace*{1cm}
    {\Large \bfseries Supplementray Information}\\[0.5cm]
    {\Large Dynamic Synaptic Modulation of LMG Qubits populations in a Bio-Inspired Quantum Brain }\\[0.3cm]
    {\large J. J. Torres$^{1,3,*}$ and E. Romera$^{2,3}$ }\\[0.5cm]
{$^1$ Departamento de Electromagnetismo y Física de la Materia and 
	Instituto Carlos I de F\'\i sica Te\'orica y Computacional, Universidad de Granada, Fuentenueva s/n, 18071 Granada,
	Spain}\\[0.5cm]
{$^2$ Departamento de F\'{\i}sica At\'omica, Molecular y Nuclear and 
	Instituto Carlos I de F\'\i sica Te\'orica y Computacional, Universidad de Granada, Fuentenueva s/n, 18071 Granada,
	Spain}\\[0.5cm]
 {$^3$ Cátedra del Consejo Social en Energía y Tecnologías Cuánticas, Universidad de Granada, 18071, Granada, España}\\[0.5cm]
    
\end{center}

\section{Hybrid digital quantum--classical implementation}

The proposed dynamics admits a natural hybrid quantum--classical digital implementation. In this scheme, the LMG evolution is performed on a quantum processor, whereas the activity-dependent synaptic feedback is evaluated by a classical controller. A schematic representation of the protocol, together with an illustrative gate-level circuit for \(N=4\) qubits, is shown in Supplementary Figure~\ref{fig:hybrid_lmg_circuit}.

We define the collective spin operators as
\begin{equation}
J_\alpha=\frac{1}{2}\sum_{i=1}^{N}\sigma_i^\alpha,
\qquad \alpha=x,y,z,
\end{equation}
where \(\sigma_i^\alpha\) denotes the Pauli operator acting on qubit \(i\). At the discrete time step \(k\), the LMG coupling is modulated by the synaptic efficacy,
\begin{equation}
g(k)=g_0 r(k).
\end{equation}
For \(h=0\), the Hamiltonian can be written, up to additive constants, as
\begin{equation}
H_{\mathrm{LMG}}(k)=-\frac{g(k)}{2N} \sum_{i<j} \left[
(1+\gamma)\sigma_i^x\sigma_j^x+
(1-\gamma)\sigma_i^y\sigma_j^y
\right ],
\end{equation}
where \(\gamma\) controls the anisotropy of the collective interaction.

The quantum layer approximates the unitary evolution over a short time interval \(\Delta t\) through a first-order Trotter decomposition,
\begin{equation}
U_k(\Delta t)
\simeq
\prod_{i<j}
U^{(ij)}_{XX}(\theta_x)
U^{(ij)}_{YY}(\theta_y),
\end{equation}
with
\begin{equation}
\theta_x=
\frac{g(k)\Delta t}{2N}(1+\gamma),
\qquad
\theta_y=
\frac{g(k)\Delta t}{2N}(1-\gamma),
\end{equation}
up to the sign and angle conventions adopted for the elementary rotation gates. Each two-qubit term is decomposed into standard one-qubit basis rotations, entangling gates, and conditional phase rotations. In the illustrative circuit of Supplementary Figure~\ref{fig:hybrid_lmg_circuit}, a SWAP-network structure is used to permute logical qubits and implement the effective all-to-all pair interactions with local-connectivity gates. When an external field \(h\neq 0\) is considered, its contribution can be incorporated through local \(R_z\) rotations associated with the \(J_z\) term.

After each Trotter block, measurements in the computational basis provide an estimate of the collective activity,
\begin{equation}
E(k)
=
\frac{1}{2}
+
\frac{\langle J_z\rangle_k}{N},
\end{equation}
which corresponds to the mean fraction of excited qubits. This value is passed to the classical controller, where the synaptic efficacy is updated according to the discrete algebraic rule
\begin{equation}
r(k+1)
=
\frac{1}{1+\tau_r \,\mathcal{U}\, E(k)}.
\end{equation}
This expression may be regarded as a quasi-stationary discrete approximation to the activity-dependent synaptic depression mechanism introduced in the continuous-time model. The updated synaptic variable then fixes the coupling strength in the next quantum step,
\begin{equation}
g(k+1)=g_0 r(k+1).
\end{equation}

The resulting protocol realizes a closed hybrid feedback loop: the quantum processor generates the collective LMG dynamics, measurements estimate the instantaneous population activity, and the classical layer updates the synaptic efficacy that modulates the subsequent quantum evolution. A detailed resource analysis of this implementation,  lies beyond the scope of the present work and will be addressed in a separate study.
\begin{figure}[ht!]
    \centering
    \includegraphics[width=\linewidth]{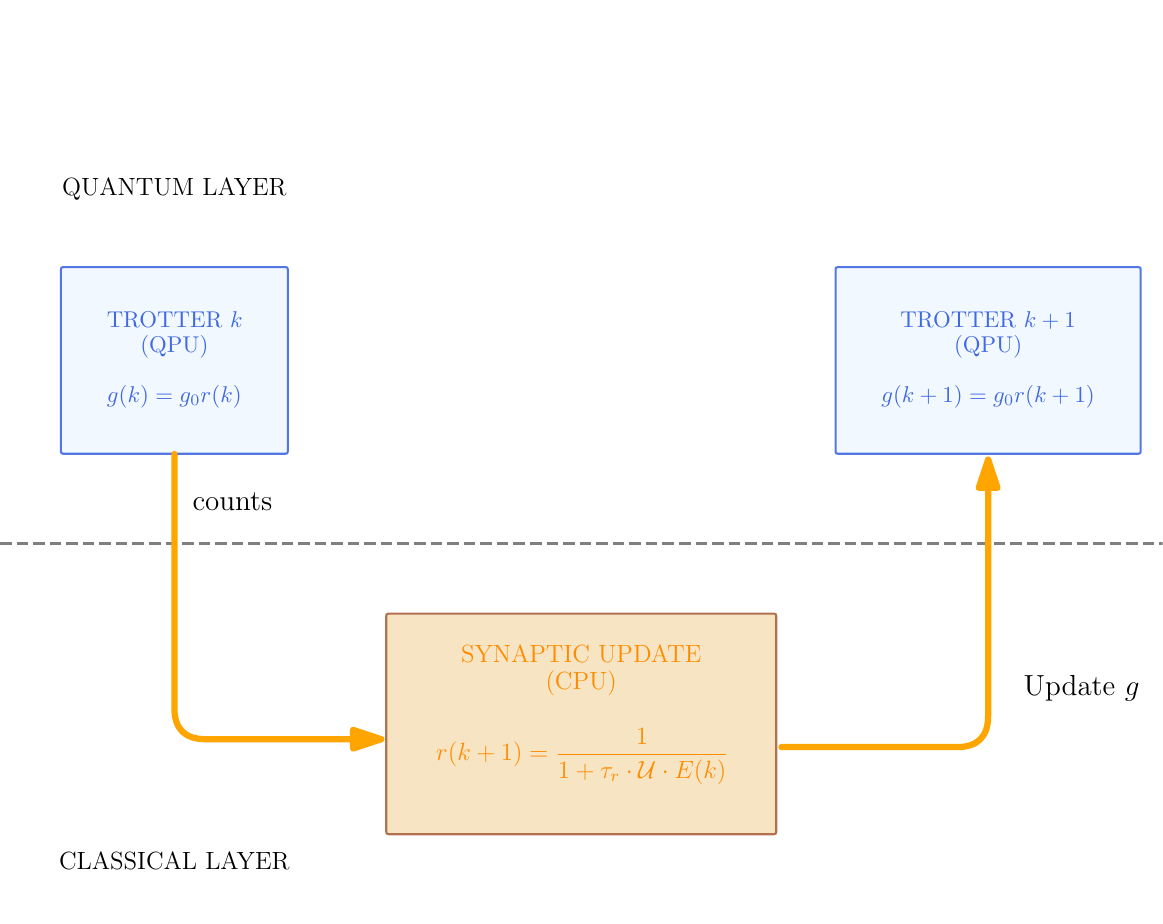}
    \vspace{0.5cm}
    \includegraphics[width=\linewidth]{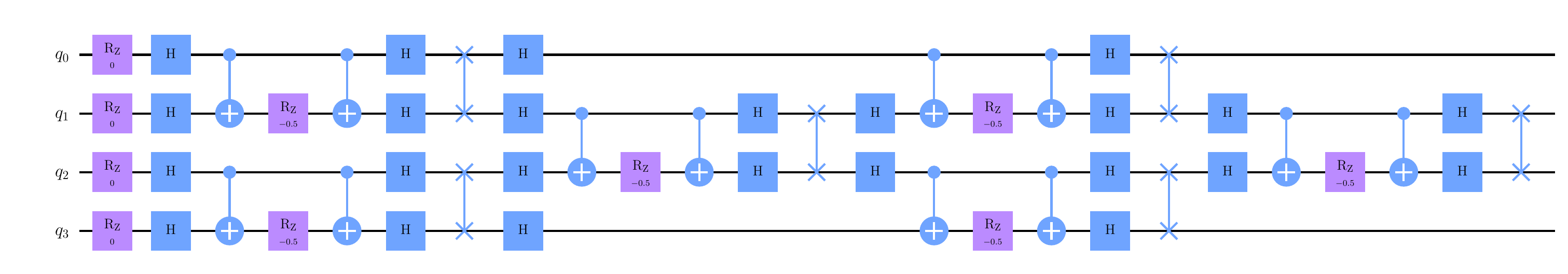}
    \caption{
    Hybrid quantum--classical implementation of the activity-dependent LMG model.
    Top: feedback protocol. The quantum layer performs a Trotterized LMG evolution with coupling \(g(k)=g_0r(k)\); measurements estimate \(E(k)=1/2+\langle J_z\rangle_k/N\), which is used classically to update \(r(k+1)\) and hence \(g(k+1)\).
    Bottom: illustrative \(N=4\) gate-level Trotter block. Pairwise interactions are decomposed into local rotations, entangling gates and conditional phase rotations, with SWAP gates enabling different qubit pairs to interact under local-connectivity constraints. 
     {For simplicity, this circuit implementation illustrates the case of $\gamma=1$. In practice, this eliminates the $Y\otimes Y$ terms, requiring only the execution of the unitary evolutions $U_{XX}^{(ij)}(\theta_x)$ between physically adjacent qubits at each layer to reconstruct the all-to-all Hamiltonian.}
    }
    \label{fig:hybrid_lmg_circuit}
\end{figure}

\section{Effect of an external magentic field $h$ }
Although the main analysis in the manuscript focuses on the case \(h=0\), the inclusion of a finite longitudinal field provides a simple way to model an external drive acting on the quantum-brain population. To illustrate the qualitative effect of this term, we considered the Hamiltonian contribution \(hJ_z\) and analyzed the resulting population dynamics for increasing values of \(h\).
Supplementary Figure~\ref{fig:external_field_dynamics} shows that even small values of \(h\) modify the interplay between the collective excitation level and the synaptic feedback variable. For weak fields, the system preserves the oscillatory homeostatic regime observed in the field-free case, with a clear anticorrelation between the collective activity and the synaptic efficacy. As \(h\) increases, the oscillations become progressively more irregular and the balance between high- and low-activity phases is altered. For sufficiently large fields, the dynamics is driven toward a strongly biased regime in which the population remains close to a highly excited state, while the synaptic efficacy is correspondingly suppressed.

These results indicate that the external-field term can act as a control parameter for tuning the collective dynamical regime of the model, ranging from homeostatic oscillations to field-dominated saturated activity. A systematic exploration of this driven regime, including its dependence on system size, anisotropy and synaptic time scales, lies beyond the scope of the present work and will be addressed elsewhere.
\begin{figure}[t]
    \centering
    \includegraphics[width=\linewidth]{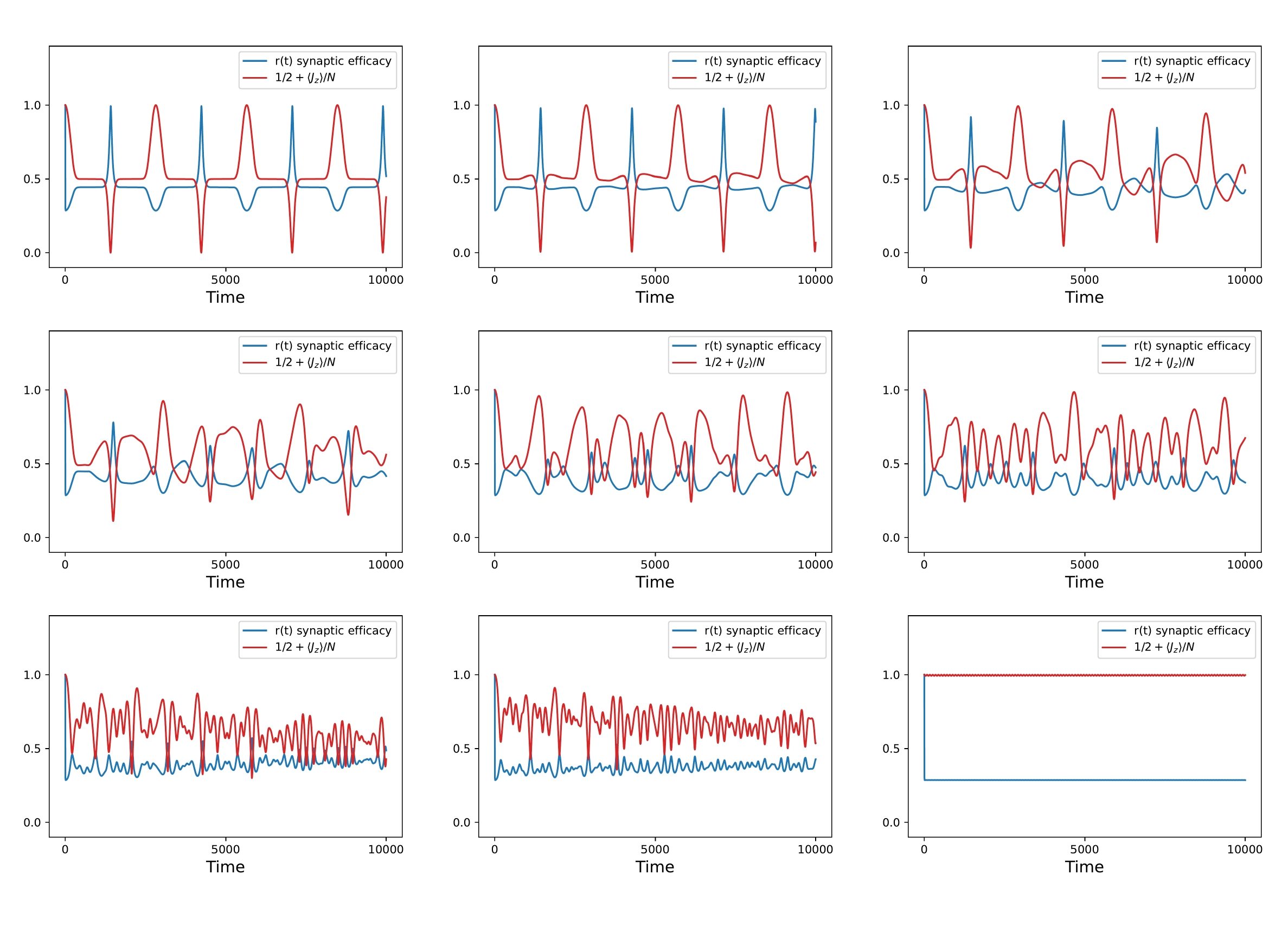}
    \caption{
    Effect of the external-field term \(hJ_z\) on the LMG quantum-brain dynamics. Each panel shows the time evolution of the collective excitation level \(E(t)=1/2+\langle J_z\rangle/N\) (red) and the synaptic efficacy \(r(t)\) (blue). From top to bottom and from left to right, the field strength is increased as
    \(h=0,\,2\times10^{-5},\,5\times10^{-5},\,10^{-4},\,5\times10^{-4},\,10^{-3},\,5\times10^{-3},\,10^{-2},\,5\times10^{-2}\).
    Other parameters are \(N=20\), \(g_0=0.05\), \(\gamma=1\), $U(t)={\cal U}=0.5$ for all time (i.e. $\tau_f=0$) and \(\tau_r=5\).}
    \label{fig:external_field_dynamics}
\end{figure}

\section{Quantifying the effect of synaptic feedback on system entanglement}
To quantify how synaptic feedback affects entanglement persistence, we analyzed the time intervals during which the entropy \(S_L(t)\) remains above a fixed threshold,
\begin{equation}
S_{\mathrm{th}}=1.40 .
\end{equation}
High-entanglement windows are defined as continuous intervals \([t_a,t_b]\) such that
\begin{equation}
S_L(t)>S_{\mathrm{th}},
\qquad
\forall\, t\in[t_a,t_b].
\end{equation}
The duration of each window is
\begin{equation}
\Delta t_i=t_b^{(i)}-t_a^{(i)},
\end{equation}
and, for each value of \(\tau_r\), the mean duration is computed as
\begin{equation}
\langle \Delta t\rangle
=
\frac{1}{M}
\sum_{i=1}^{M}\Delta t_i ,
\end{equation}
where \(M\) is the number of detected high-entanglement windows.

Supplementary Figure~\ref{fig:entanglement_persistence} shows that \(\langle \Delta t\rangle\) increases with \(\tau_r\), indicating that slower synaptic recovery prolongs the residence time of the system in highly entangled states. Thus, synaptic feedback modulates not only the collective activity dynamics, but also the temporal persistence of entanglement.
\begin{figure}[ht!]
    \centering
    \includegraphics[width=\linewidth]{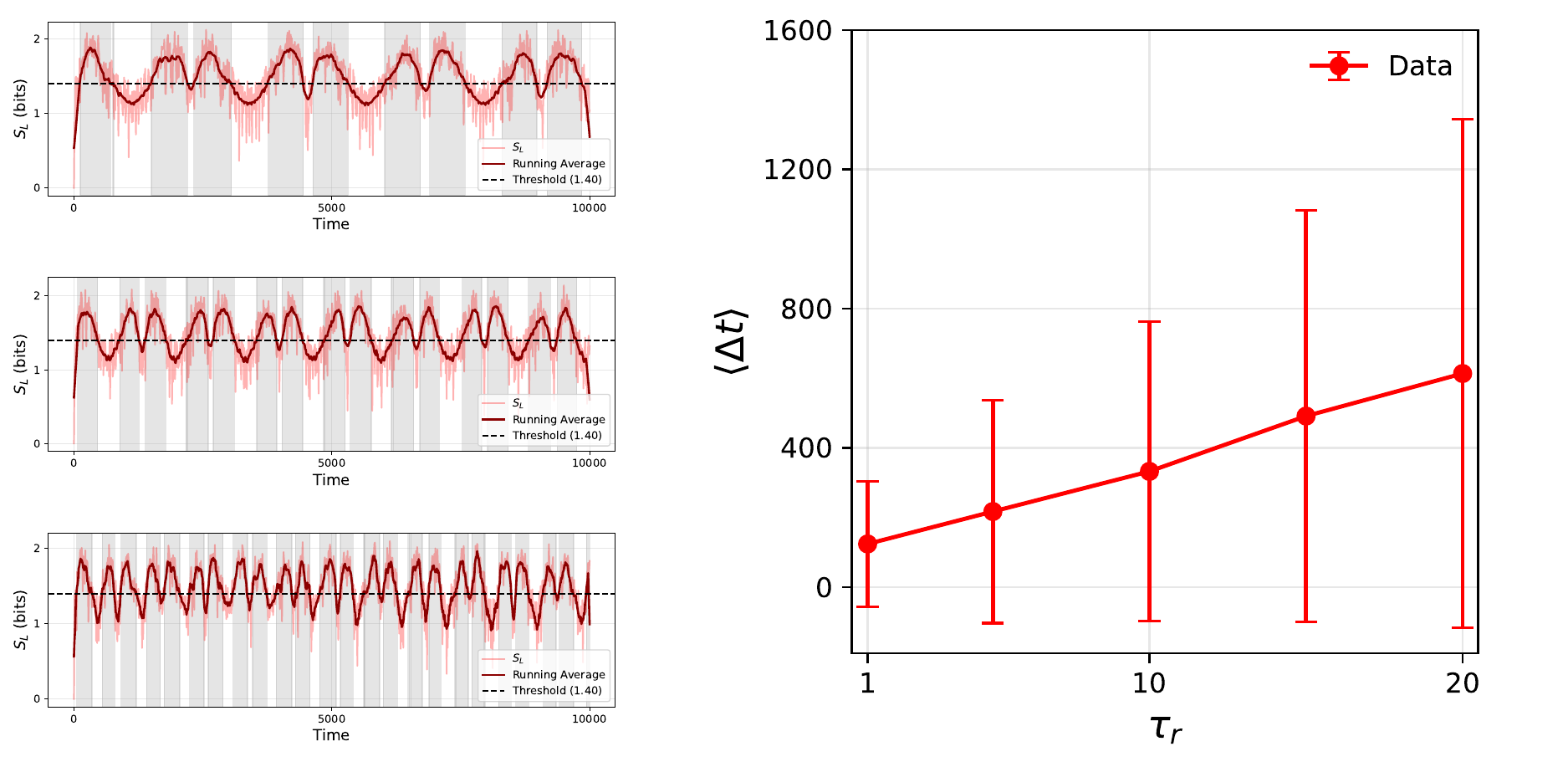}
    \caption{
    Entanglement persistence as a function of synaptic feedback. Left: temporal evolution of the von Neumann entropy $S_L(t)$   for three increasing levels of synaptic depression being
$\tau_r = 0, 10, 20$ from top to bottom, respectively, running average (red solid line), and threshold \(S_{\mathrm{th}}=1.40\) (black dashed line). The grey areas show the time windows $\Delta t$ for high-entanglement. Right: mean duration \(\langle \Delta t\rangle\) of the high-entanglement windows as a function of \(\tau_r\). Other parameters were $N=20,$ $g_0=5,$ $\gamma=0.8,$ $U(t)={\cal U}=0.5$ for all time (i.e. $\tau_f=0$) and $r_0=1.$}
    \label{fig:entanglement_persistence}
\end{figure}

\end{document}